\documentclass[%
 reprint,
 amsmath,amssymb,
 aps,
]{revtex4-2}

\usepackage{braket}
\usepackage{xspace}
\usepackage{tikz,quantikz}
\usepackage{orcidlink}
\usepackage{comment}
\usetikzlibrary{external}
\usepackage{amssymb}
\usepackage[T1]{fontenc}

\usepackage{import}
\usepackage{transparent}

\usepackage[mode=build]{standalone}
\newcommand{\ea}{\textit{et al.}\xspace}

\usepackage{graphicx}
\usepackage{dcolumn}
\usepackage{bm}

\begin{document}

\preprint{APS/123-QED}
\title{Flexible quantum data bus for quantum networks}

\author{Julia Freund\orcidlink{0000-0001-5548-5007}$^{1}$, Alexander Pirker$^{1}$ and Wolfgang Dür\orcidlink{0000-0002-0234-7425}$^1$}
 \affiliation{$^1$ Universität Innsbruck, Institut für Theoretische Physik, Technikerstraße 21a, 6020 Innsbruck, Austria}

\date{\today}

\begin{abstract}
We consider multipath generation of Bell states in quantum networks, where a preprepared multipartite
entangled two-dimensional cluster state serves as a resource to perform different tasks on demand. We show
how to achieve parallel connections between multiple, freely chosen groups of parties by performing appropriate
local measurements along a diagonal, staircase-shaped path on a two-dimensional cluster state. Remarkably, our
measurement scheme preserves the entanglement structure of the cluster state such that the remaining state is
again a two-dimensional cluster state. We demonstrate strategies for generating crossing, turning, and merging
of multiple measurement lines along the two-dimensional cluster state. The results apply to local area as well as
to long-distance networks.
\end{abstract}

\maketitle

\section{Introduction}Connecting devices constitutes the foundation of modern society. All devices operate in some form of network, with the goal of either increasing computational power, to perform distributed tasks or to simply exchange messages of some kind. The structure and implementations of such networks often depend on the sizes of devices. For example, for embedded devices such a network corresponds to connected, hard-wired microcontrollers. In contrast, computers connect through local-area-networks over Wi-Fi. With the advent of quantum computers approaching, researchers believe that connecting quantum devices via quantum networks is a crucial step to unlock new applications which do not have a classical counterpart~\cite{Cacciapuoti18,Cacciapuoti2020}. Such applications include quantum key distribution~\cite{GisinRevMP2002}, secret sharing~\cite{Hillery99,Markham08}, sensor networks~\cite{Eldredge2018,SekatskiPRR2020} but also distributed quantum computing~\cite{DistQuantumComputingBuhrman1997PRA,DistributedQuComputing,caleffi2022distributed}. 

At the lowest level, quantum devices, in any quantum network~\cite{QuantumInternetKimble,QuantumInternetWehner,Azuma2021,Azuma2023,Illiano2022}, at any scale, are connected via physical channels. A Bell state between nodes is equivalent to a channel due to the teleportation protocol~\cite{BennettTele}, paving the way for sharing and manipulating entanglement~\cite{Chen2023} as an alternative to channels. In recent years, the vast majority of research on quantum networks has focused on the generation of Bell states between two nodes in a network of Bell states~\cite{Meter2013b,Pirandola2019,Liorni_2021,LIIEEE2021}, the routing of quantum information carriers~\cite{Schoute2016,Caleffi17,Viscardi2023MarkovProcess,Gyongyosi2017,Gyongyosi18,Das17} in quantum networks. The generation of Bell states involves finding an optimal path~\cite{Caleffi17} (in terms of some metrics) within the network and connecting the states along that path. Here, we introduce a flexible quantum data bus by establishing multiple Bell states simultaneously in a network of any scale by utilizing pre-established two-dimensional (2D) cluster states as a resource, referred to as cluster states~\cite{BriegelPRL2001Cluster,RaussendorfPRA2003MBQCCluster} in the following, instead of a network of Bell states as a resource for routing~\cite{Pirker_2018,Pirker_2019}.

In our setting, we assume that each network node stores only a single qubit. The nodes share a two-dimensional cluster state, which has a grid structure, and interconnects the nodes and devices, respectively. This grid-state serves as a flexible network state to fulfill different tasks on demand, and can be established even prior to network requests. The manipulation of network states takes place solely by single-qubit measurements and single-qubit unitaries. This falls within the general spirit of entanglement-based quantum networks with preshared entanglement~\cite{Pirker_2018,Pirker_2019}. Such an approach offers not only high flexibility~\cite{MiguelRamiro2023optimizedquantum}, but also avoids latencies.

In this work we report the following two main results:
\begin{enumerate}
\item[(i)] A zipper scheme to establish a connection between two nodes within a two-dimensional cluster state, which restores the structure of the remaining two-dimensional cluster state.
\item[(ii)] A quantum data bus that allows to generate multiple Bell states simultaneously by measuring along parallel measurement lines along the two-dimensional cluster state.
\end{enumerate}

At the heart of our proposal is the zipper scheme (i), a measurement pattern among diagonal lines to establish a Bell state within the grid of the two-dimensional cluster state. We remark that the zipper scheme is based on the X protocol introduced in Ref.~\cite{Hahn2019}. Other works also study staircase-shaped measurement patterns on 2D cluster states, even demonstrating optimality with regards to the number of measurements \cite{morruiz2023influence,mannalathPRAmultiparty}. As we demonstrate, the zipper scheme preserves the entanglement structure of the remaining cluster state when generating a Bell state. Keeping the grid structure intact and usable turns out to be pivotal to extract further Bell states from the cluster state. We utilize this remarkable property to establish multiple Bell states between arbitrary nodes in a cluster state, including crossings and turns of the measurement paths. We note, however, that there exist some prohibited areas around the turning and endpoints, where Pauli $Z$ measurements are necessary to cut out the Bell state from its direct neighborhood. However, this is still in stark contrast to standard methods~\cite{RaussendorfOneWayPRL2001, RaussendorfPRA2003MBQCCluster,Briegel2009}, where the connection between nodes is established by cutting a hole among the entire path. This method not only wastes resources, but also hinders the generation of further Bell states. Our scheme enables to simultaneously create Bell states along adjacent measurements paths, effectively constructing a quantum data bus (ii) in analogy to classical buses found for instance in mainboards. We compose a quantum data bus architecture in a modular fashion by combining elementary building blocks such as crossings, turns and merging or splitting of multiqubit measurement lines in a cluster state. Our results have a multitude of applications from which we discuss its applicability ranging from large-scale networks such as entanglement-based quantum networks~\cite{Pirker_2018,Pirker_2019,Meignant2019,Epping_2016} to local networks relying on a central unit~\cite{Cuquet2012,Avis2023}. Furthermore, we believe that also small and integrated networks of quantum devices, similar to devices running in embedded environments like cars, benefit from a dynamic way of creating Bell states on demand, as this removes the necessity to multiplex multiple connections over a single transmission channel like in classical, hardwired controller networks. Diagonal routing also naturally extend to the multiparty case, like for example establishing Greenberger-Horne-Zeilinger (GHZ) states between multiple nodes within the cluster.

This work is organized as follows. In section~\ref{sec:setting} we introduce our network setting based on entangled network states. In particular, we utilize graph states and their local manipulations to obtain the network requests, as we explain in section~\ref{sec:Preliminaries}. Subsequently, we present our main finding, the zipper scheme in section~\ref{sec:zipper}, and show how we can use these result to build a modular quantum data bus in section~\ref{sec:bus}. In section~\ref{sec:apps} we focus on three applications of quantum networks which profit from our findings. Finally, we conclude our work and point out future research directions in section~\ref{sec:conclusion}.

\section{Setting}\label{sec:setting}
We consider quantum networks at different scales with the goal of establishing entanglement between multiple pairs of nodes in parallel. Rather than generating Bell pairs directly, we consider an entanglement-based network~\cite{Pirker_2018,Pirker_2019} where nodes share a preprepared multipartite entangled network state, which is generated during idle times of the network. In this work we suggest a flat hierarchy among the nodes. In particular, we assume that each node holds one qubit of the cluster state, and thus all nodes have the same functionality and privileges. The nodes manipulate the network state solely by local unitaries and single-qubit measurements to establish the desired target state in a flexible way on demand. This approach does not require to send any quantum information carriers when the network request arrives and consequently reduces the latency to serve requests. Moreover, the network state can be optimized with respect to the connectivity demands of the network~\cite{MiguelRamiro2023optimizedquantum,morruiz2024imperfect} such as adjusting to local connectivity demands of client groups. On the contrary entanglement-based networks require a long time memory for storage. Here we consider a 2D cluster state as a resource, where each node holds a single qubit of the cluster state. Such 2D cluster states serve as universal resource in measurement-based quantum computation (MBQC)~\cite{RaussendorfOneWayPRL2001, RaussendorfPRA2003MBQCCluster,Mantri2017}, where information is processed by means of single-qubit measurements only and arbitrary target states can be generated in this way. Here, in contrast, our sole goal is to establish multiple Bell pairs, which implies that considerations such as information flow~\cite{Danos2006,Browne_2007} do not apply, and additional methods and techniques are available. In particular, we are interested in establishing (multiple) Bell pairs between arbitrary nodes, independent of their location within the cluster state.

\section{Preliminaries}\label{sec:Preliminaries}
Graph states~\cite{heinEntanglementGraphstates,HeinPRA2004} are pure, multipartite entangled quantum states described by a classical graph $G=(V,E)$, where the vertex set $V$ and edge set $E$ correspond to qubits and binary Ising-type interaction between those, respectively. Formally, a graph state $\ket{G}$ is uniquely defined as the eigenvector with $+1$ eigenvalues of the set of stabilizers
\begin{equation}
K_a=\sigma_x^a \sigma_z^{N_a}=\sigma_x^a \prod_{b \in N_a} \sigma_z^b
\end{equation}
for each vertex $a \in V$ with its corresponding neighbors $N_a \subseteq V$, and note that $\sigma_i$ denotes the $i$-th Pauli matrix. The dynamic picture is an alternative means to define graph states, where all qubits are initially prepared in the $\ket{+}$ state and according to the edge set controlled-phase gates are applied between two vertices.

In this work we frequently use local complementation (LC), a specific local Clifford unitary, to transform a graph state. This unitary inverts the subgraph induced by the neighbors of a qubit. Local Pauli $Z$ measurements translate to vertex and edge erasure of the measured qubit. A Pauli $Y$ measurement corresponds to a LC followed by a Pauli $Z$ measurement on the qubit to measure. Local Pauli $X$ measurements are a combination of LC on a specific neighbor qubit, a Pauli $Y$ measurement on the qubit itself and again LC on the specific neighbor, see Ref.~\cite{HeinPRA2004}.

\section{Zipper scheme}\label{sec:zipper} 
\begin{figure}
    \includegraphics[]{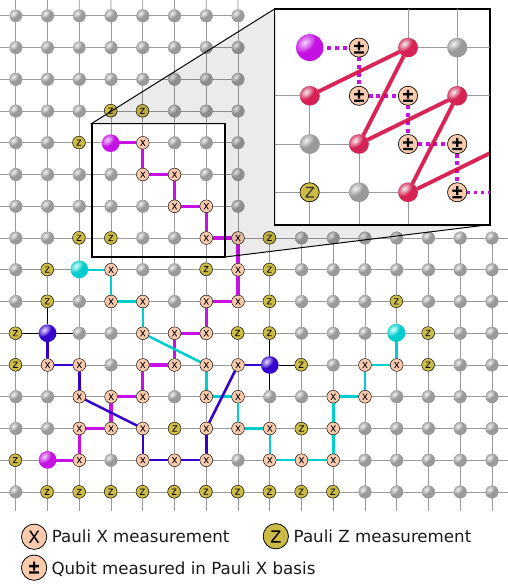}
\caption{\label{fig:zipperScheme} Zipper scheme (inset): The orange qubits are measured in the Pauli $X$ basis such that the purple qubit is routed along the diagonal measurement path while the obtained edges between the red qubits restore the underlying cluster state. The qubits in the neighborhood of the purple qubit need to be removed by Pauli $Z$ measurements leading to holes. Simultaneous Bell state routing (main figure): We extract three Bell states, purple, blue and turquoise, by applying the zipper scheme along diagonals of orange qubits. Merely on turning and endpoints holes appear due to the yellow Pauli $Z$ measurements to disconnect the Bell state qubits from the remaining cluster state. The zipper scheme enables crossing of paths except in the vicinity of end and turning points, and note that the paths are adjusted such that the measurement sequence is taken into account (first purple, then turquoise and blue).}
\end{figure}

The zipper scheme tackles the problem of connecting two nodes in a 2D cluster state via two diagonal lines, thereby generating a Bell state. The scheme accomplishes this by performing Pauli $X$ measurements along a staircase-shaped measurement path (as introduced in~\cite{Hahn2019} as $X$ protocol), where two such paths are combined to connect any two points on the grid. For a direct diagonal connection, one of the Bell state qubits is used as neighbor for LC, and Pauli $Z$ measurements in the neighborhood of the endpoints isolate the final Bell state. The inset in Fig.~\ref{fig:zipperScheme} illustrates the result of the zipper scheme after the orange qubits have been measured in the $X$ basis along a diagonal, whereas the purple qubit is one of the Bell state constituents and has been used as reference qubit for LC. We observe that measuring the orange qubits along the diagonal, purple path introduces the red edges between the red qubits on each side of the measurement path, these edges restore the underlying 2D cluster state structure. 
However, some holes are caused by Pauli $Z$ measurements in the direct neighborhood of end- and turning points, which are necessary to fully isolate the Bell state, the yellow qubits in Fig.~\ref{fig:zipperScheme} correspond to this measurements.

The zipper scheme allows us to generate a Bell state on a diagonal line within the cluster state. However, we observe, as explained in more details in the Appendix (note that routing a single Bell state is a special case of routing parallel Bell states), that it is also possible to use the very same scheme to route a Bell state around corners (\textquotesingle L\textquotesingle -turn), upside down (\textquotesingle V\textquotesingle -turn), diagonal-to-straight or over a crossing with another Bell state. These ingredients enable us to route measurement lines for Bell states of arbitrary topology on the cluster state. In Fig.~\ref{fig:zipperScheme} we demonstrate an example in which three Bell states among arbitrary nodes are generated. We emphasize that two points on a 2D lattice can be always connected by two (or more) diagonals as shown in the Fig.~\ref{fig:zipperScheme}. Importantly, the zipper scheme allows us to establish crossings of Bell states on the cluster state provided that the crossings do not involve on qubits that are direct neighbors (yellow qubits to be measured in $Z$ basis) of the end or turning points.

Note that we do not sketch the additional edges after applying the zipper scheme to simplify readability, and furthermore, we have adjusted the measurement paths assuming that the purple path is measured first followed by the turquoise and blue one. The ability to support crossings has two advantages. First, it prevents potentially long measurement paths around obstacles in a cluster state, and second, it allows to have simultaneous request also in cases that require crossings due to resource constraints, impossible to achieve without them. 
Using a 2D cluster state of size $n \times n$ allows one to obtain $\mathcal{O}(n)$ parallel Bell pairs. This can be seen by observing that any two points in a 2D cluster are connected by diagonal paths of length at most $2n$, which also corresponds (up to few additional measurements around end- and turning points) to the number of qubits that need to be measured to establish a Bell pair. Since the underlying entanglement structure is still intact, in total $\mathcal{O}(n)$ Bell pairs can be generated generically. Notice that some configurations may not be possible, as small holes appear in the two-dimensional cluster state from $Z$ measurements of qubits that directly neighbor end- and turning point qubits on the measurement path.

\section{Quantum data bus} \label{sec:bus}
\begin{figure}[hb]
    \centering
   \includegraphics{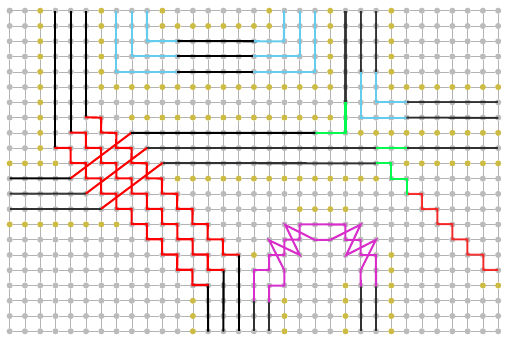}
    \caption{Quantum data bus and its building blocks: The red lines represent parallel transport along diagonal lines, including crossings orthogonal to the diagonals. The black lines show a parallel transport, both vertical and horizontal. The blue and pink lines depict parallel \textquotesingle L\textquotesingle - and \textquotesingle V\textquotesingle -turns, whereas the green lines demonstrate a merging or splitting of data lines. Finally, the yellow qubits depict qubits on which we need to perform Pauli $Z$ measurements.}
    \label{fig:Databus}
\end{figure}

The next step is to extend the results of the zipper scheme and building blocks to generate multiple Bell states along parallel lines, which we refer to as a quantum data bus. We note that a quantum data bus is conceptually closely related to data buses in conventional computers used for connecting individual microchips with each other. First, our scheme generates multiple, spatially separated Bell pairs along parallel measurement lines, which is analogously found in classical data buses. The second main feature of a classical data bus is that it is bidirectional, which means that information can be sent in both directions, and our measurement scheme allows for that. A quantum data bus uses the entanglement of a cluster state in an efficient way to achieve that goal. In Fig.~\ref{fig:Databus} we illustrate the main idea.

Similar when generating a single Bell state along a measurement line, we identify multiple building blocks for generating Bell states in parallel. We point out that the building blocks we discussed in the previous section extend naturally to parallel building blocks. Foremost, is the diagonal crossing of Bell states, which follows trivially from the state restoring properties of the zipper scheme, and we represent it by the red structure in Fig.~\ref{fig:Databus}. For instance, from a cluster state of size $n\times n$, it is possible to generate $\mathcal{O}(n)$ parallel Bell states as part of a crossing. The entanglement structure remains when one applies the zipper scheme, and this key feature gives rise for further building blocks such as the \textquotesingle L\textquotesingle (blue blocks) and \textquotesingle V\textquotesingle -turn (purple blocks) as well as the straight line measurement lines (black blocks), which we discuss partly here and in the Appendix~\ref{app:V}.

\begin{figure}[ht]
    \centering
    \includegraphics[trim={0cm 20cm 0 0},clip]{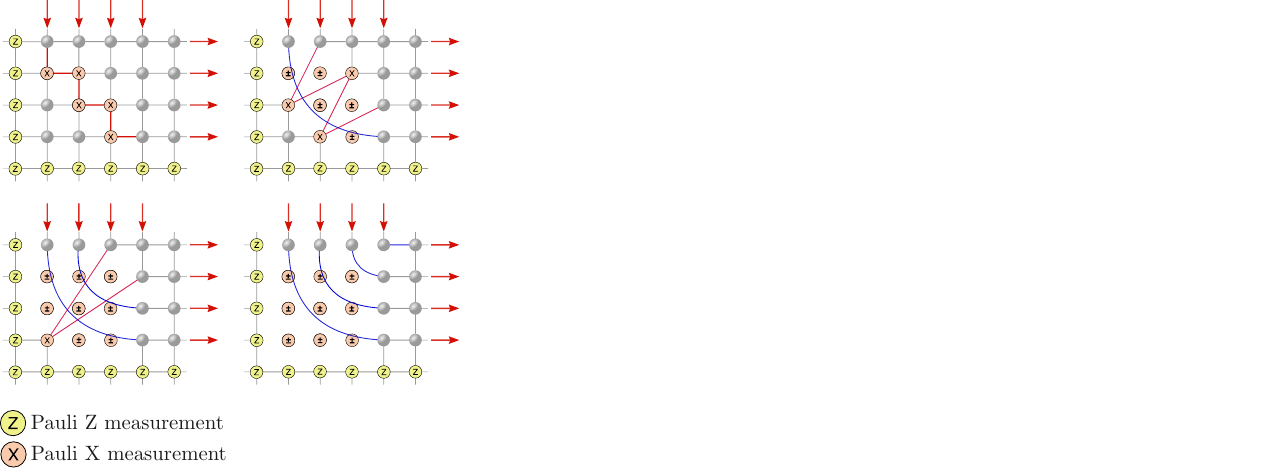}
    \caption{The figure shows a step-by-step guide for the measurement pattern to perform a perpendicular turn from vertical to horizontal in a 2D cluster state with four parallel data lines. The zipper scheme is applied from the left- to right-most qubits, where each application of the zipper scheme enables the next measurement. Note that the red arrows are only visual elements to better see the turn of data lines, and the arrows can be also reversed.}
    \label{fig:corner}
\end{figure}

Here we begin with the \textquotesingle L\textquotesingle -turn building block for the quantum data bus, which can be used to change the direction of the measurement lines from vertical to horizontal (and vice versa), which illustrate step-by-step in Fig.~\ref{fig:corner}. We apply the zipper scheme (in inward direction) to the outermost data line until we diagonally reach the inner most data line, resulting in a perpendicular turn as illustrated by the leftmost blue line of Fig.~\ref{fig:corner}. The cluster state preservation property of this first step enables us to apply the zipper scheme to turn the second most left qubit as shown in the third step of Fig.~\ref{fig:corner}. Repeatedly applying this pattern results in a full turn of all data lines, as illustrated in the last step of Fig.~\ref{fig:corner}. Note that the required number of qubits to turn $n$ data lines in an \textquotesingle L\textquotesingle -turn around the corner is $n\times n$.

A new building block corresponds to parallel transport among a straight line. We achieve this kind of transport by applying the zipper scheme repeatedly. We point out that one can understand this building block simply from the previous one, if we extend the zipper scheme in Fig.~\ref{fig:corner} to the yellow left most qubits by $X$ measuring the top qubits (indicated by the arrows). Specifically, first one applies the zipper scheme from the first Bell state line towards the last line. This effectively routes the first Bell state diagonally to the last line. Next, one applies the same scheme to the second Bell state, moving it to the second last line. Repeatedly applying this pattern results in a total permutation of all lines. Performing the same scheme another time inverts the permutation, resulting in straight lines again. However, it should be noted that additional Pauli $Z$ measurements are necessary to isolate the established data bus as shown by yellow dots and in Fig.~\ref{fig:Databus} and Fig.~\ref{fig:corner}. A single vertical qubit line is necessary to separate the two permutation measurements, because the zipper scheme requires be applied on a complete a cluster state. In total it requires a block of length $2n$ qubits to transport $n$ data lines in parallel, which induces a fixed, minimal block size.
Notice, although that there is no distance between the data lines within the data bus, in stark contrast to standard data transport schemes in MBQC \cite{RaussendorfOneWayPRL2001,RaussendorfPRA2003MBQCCluster,Briegel2009} 
which require to isolate a path, resulting in a distance of one between two data lines, i.e. half the capacity as in our scheme.

Another new building block allows us to merge or split data lines to or from the data bus. In Fig.~\ref{fig:Databus} the green structure represents a splitting of data lines, and it uses Pauli $Z$ (yellow dots) and $X$ measurements (green dots) to separate and merge the data lines, respectively. For adding a single data line we require one Pauli $Z$ and one Pauli $X$ measurement to cut and merge the line into the data bus.

Similar approaches as presented apply for other building blocks as well, we refer to the Appendixes~\ref{sec:appendix:parallel},~\ref{app:V}, and \ref{app:merging} for details. For \textquotesingle V\textquotesingle -turns and the building block that turns diagonal Bell states into a parallel Bell states, the number of data lines $n$ of the building block determines the minimum, necessary length of the zipper scheme on the first data line as $2n$. Therefore, it requires in total $n\times n$ measurements. We also point out that diagonal routing straightforwardly extends to GHZ states, as described in the Appendix~\ref{app:ghz}. Essentially, one can connect several qubits to some central one via different (diagonal) paths, thereby establishing a GHZ state.

The size of the individual building blocks allows one to estimate the size of the required cluster state depending on the desired information transport. 

\section{Applications} \label{sec:apps} We have laid the theoretical foundation for routing Bell states on a 2D cluster state. In the following we discuss three potential use-cases of the parallel zipper scheme, highlighting its applicability and importance. 

The first application corresponds to a long-distance quantum network in which the nodes, which hold a single qubit each, connect via a two-dimensional cluster state. Note that we allow these nodes to be simple clients as well as sophisticated quantum local area networks (LAN) nodes with internal structure. This setting was studied, e.g., in \cite{Wallnofer2016,Wallnofer2019}, in which the authors presented ways how to create a long-distance cluster state by purifying and merging small buildings blocks. Once the cluster state is established, the goal of the network is to generate Bell states between any of the nodes in the cluster state on demand, maybe even in parallel. Using the results of this work we find that it is possible to connect any, sufficiently far, sets of communication partners via Bell states by using the parallel zipper scheme.

But also centralized communication scenarios with a powerful central unit as studied in \cite{Cuquet2012,Avis2023} benefit from our results. For that purpose, suppose that the nodes of a small scale network, similar to a LAN, connect to a central unit via (potentially multiple) Bell states. In case the central unit uses internally a local cluster state, and the parties connect via a Bell state to the border of the cluster state, the central unit is able to establish arbitrary, even parallel, connections among the nodes by using our quantum data bus proposal. In other words, the central unit uses the internal cluster to permute connections among the (outside) nodes. This allows for a flexible way to operate the central unit, as this unit is capable of connecting any of the connected nodes as necessary and on demand. The advantage of having a pre-established cluster state over distributing Bell pairs is that each node only needs a single quantum memory at each node. Moreover, this method required less latency compared to distributing the Bell state on demand.

The last application we highlight corresponds to integrated quantum devices. In this vision, small quantum devices like sensors connect  within the integrated quantum device via a Bell state to the border of a 2D cluster state, similar in spirit to the previous application. This scenario is inspired by classical computing architectures in which microcontrollers connect through buses in a fixed manner. Our results imply that for quantum computing architectures such a fixed wiring is not necessary, but can be established on-demand in terms of quantum data buses by consuming a cluster state. This introduces flexibility, but also extendability to integrated quantum device design.

\section{Conclusion}\label{sec:conclusion} In this paper we have considered pre-established 2D cluster states as resource for quantum networks. We identified the main ingredient to route Bell states in a cluster state, namely the zipper scheme. The scheme preserves the cluster state entanglement structure, vital for establishing crossings of Bell states on a cluster. It further turns out that the zipper scheme lies at the heart of many building blocks to route Bell states in a cluster state, such as, for example \textquotesingle V\textquotesingle -turns or \textquotesingle L\textquotesingle -turns. We also demonstrated how to further optimize these building blocks by showing how to utilize them for parallel transport of Bell states. These building blocks play a key role for routing multiple Bell states in a long-distance network, but also for small-scale networks and the design of integrated quantum processing devices. We also discussed an extension of the diagonal routing to multiparty states, such as, for example, GHZ states, and we leave the extension of the other building blocks for future work. As an outlook, it remains an open question how noise in the network state and the apparatus influences the fidelity of the routed Bell states. The first results for a single target Bell pair \cite{morruiz2023influence} indicate robustness against imperfections. 

\section{Acknowledgments}
This research was funded in whole or in part by the Austrian Science Fund (FWF) 10.55776/P36010. For open access purposes, the author has applied a CC BY public copyright license to any author accepted manuscript version arising from this submission.

\appendix

\section{Graph states}
\begin{figure}[ht]
    \includegraphics[]{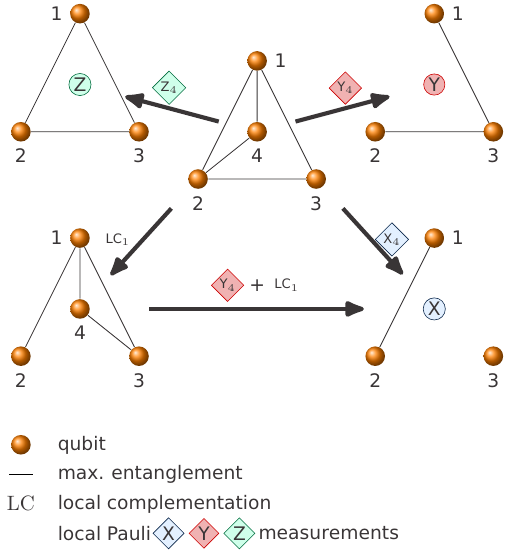}
    \caption{\label{fig:graphManipulations} In the center is a graph state depicted, whereas vertices correspond to qubits, which are initialized in the $\ket{+}$ state and edges represent maximally entanglement between two qubits. In the corners the resulting graph states are shown, which are caused by local single-qubit Pauli measurements or local complementation as indicated by the labeled arrows.}
    \end{figure}

Graph states~\cite{heinEntanglementGraphstates,HeinPRA2004,BriegelPRL2001Cluster} are a specific class of pure, multipartite entangled quantum states whose correlations correspond to a classical graph $G=(V,E)$. The vertex set $V$ and the edge set $E$ correspond to qubits and binary Ising-type interaction between qubits, respectively. Fig.~\ref{fig:graphManipulations} depicts an example graph state in the center. The figure demonstrates how the graph state transforms under certain local operations, like for example local complementations or Pauli basis measurements. Formally, a graph state $\ket{G}$ with respect to the graph $G$ is constructed from the set of qubits in the $\ket{+}$ state through the interaction
\begin{equation}
\ket{G}=\prod_{\{a, b\} \in E} CZ_{a b}\ket{+}^{\otimes |V|},
\end{equation}
where a controlled-$Z$ gate defined as $diag(1,1,1,-1)$ is applied, if the edge between the qubits $a$ and $b$ exists in the graph $G$. For example the central graph in Fig.~\ref{fig:graphManipulations} comprises four qubits, labeled from one to four. One generates the corresponding graph state by applying controlled-$Z$ gates between the qubits one and two, one and four, one and three, two and three, and two and four, respectively. 

Alternatively, a graph state $\ket{G}$ is defined as the common, unique eigenstate with eigenvalue $+1$ of the set of stabilizers
\begin{equation}
K_a=\sigma_x^a \sigma_z^{N_a}=\sigma_x^a \prod_{b \in N_a} \sigma_z^b, \label{eq:stab}
\end{equation}
for each vertex $a$ with its corresponding neighbors $N_a$. Consequently, we describe the central graph in Fig.~\ref{fig:graphManipulations} uniquely via the stabilizers $K_1=\sigma_x^1 \sigma_z^2 \sigma_z^3 \sigma_z^4$ and $K_2=\sigma_x^2 \sigma_z^1 \sigma_z^3 \sigma_z^4$, $K_3=\sigma_x^3 \sigma_z^1 \sigma_z^3$ and $K_4=\sigma_x^4 \sigma_z^1 \sigma_z^2$.

\subsection{Local Clifford unitaries and Pauli measurements}\label{app:sec:LCUPM}
Local transformations of graph states play a crucial role in distributed settings. Many of them correspond to graphical manipulations rules on the classical graph $G$, including local, single-qubit Clifford operations and Pauli measurements. For example, local complementation corresponds to applying the unitary operator $\sqrt{K_a}$ to the graph state $\ket{G}$. Graphically, a local complementation on vertex $a$ inverts the subgraph on the neighbors of $a$. The lower left corner of Fig.~\ref{fig:graphManipulations} demonstrates a local complementation at qubit one, inverting the subgraph consisting of the qubits two, three and four. A $\sigma_z$ measurement on qubit $a$ is equivalent to deleting all edges incident to qubit $a$, and a $\sigma_y$ measurement on qubit $a$ corresponds to a local complementation followed by a $\sigma_z$ measurement on qubit $a$. Finally, a $\sigma_x$ measurement on qubit $a$ is implemented by performing a local complementation on a neighbor $b \in N_a$, measuring $a$ in $\sigma_y$ basis and performing again a local complementation on $b$. In Fig.~\ref{fig:graphManipulations} possible local Pauli basis measurements on a graph state and their results are summarized.

\subsection{State extraction methods}
It was shown in \cite{DahlbergNPcomplete} that it is in general NP-complete to decide if a graph state can be extracted from a given graph state by local Clifford operations, local Pauli measurements and classical communication. Several methods have however been discussed to extract specific target states from resource states in the past. We review two approaches to extract single Bell states from a given resource state, where we refer to the first one as the isolation strategy and the second one as the X protocol~\cite{Hahn2019}.

\subsubsection{Isolation strategy}
The idea of the isolation strategy boils down to the isolation of the shortest path between the Bell state constituents by performing Pauli $Z$ basis measurements on the neighbors. To establish the Bell state between the end vertices one measures all intermediate qubits along the isolated path using Pauli $Y$ or Pauli $X$ measurements. This strategy has been proposed by Raussendorf \ea \cite{RaussendorfOneWayPRL2001} to isolate (circuit) wires in a measurement-based quantum computer. This approach comes with several disadvantages. First, it necessitates numerous measurements in the Pauli $Z$ basis to isolate the path, introducing noise to the cluster state. Second, it fails to preserve connectivity (in terms of entanglement) within the cluster, as Pauli $Z$ measurements essentially cut holes into the cluster state. The latter might be problematic, especially if the goal of a protocol corresponds to having multiple, parallel Bell states.

\subsubsection{X protocol}
An alternative protocol to extract Bell states from a cluster state has been proposed by Hahn \ea \cite{Hahn2019}, referred to as X protocol. It first measures all qubits along the shortest path in the Pauli $X$ basis and then removes all qubits still adjacent to the Bell state by Pauli $Z$ measurements. Ref. \cite{Hahn2019} shows that the X protocol requires in the worst case the same number of measurements as the isolation strategy. Note that the butterfly network \cite{butterfly} is the smallest example for the X protocol. Other works \cite{mannalathPRAmultiparty,morruiz2023influence} have shown that a staircase-formed measurement path should be preferred over a straight path in a 2D cluster state, because it reduces the remaining Pauli $Z$ basis measurements to the initial neighbors attached to the Bell pair. In the following section we report the potential of the X protocol for establishing multiplexed communication in a two dimensional cluster state.

\section{Proof of zipper scheme}\label{app:zipper}
\begin{figure*}[ht]    \centering
    \includegraphics[trim={0cm 23.3cm 0 0},clip]{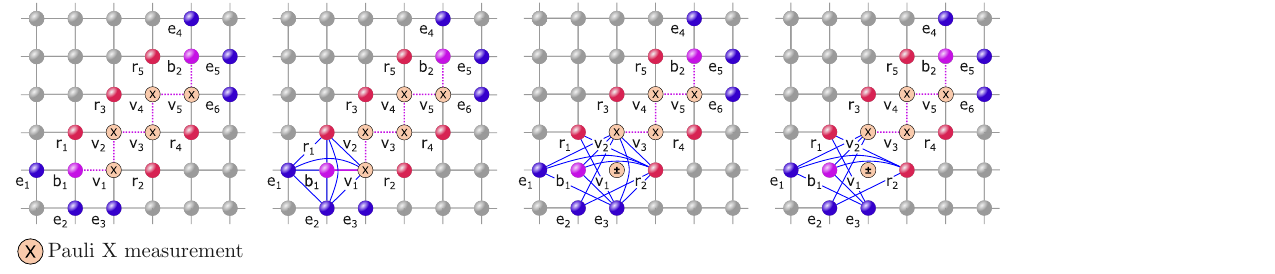}
    \caption{Core of zipper scheme: The aim is to establish the Bell-pair $(b_1,b_2)$ with the zipper scheme on the diagonal $v_1$ to $v_6$, the initial setting is shown in first step. The second step shows the result of a local complementation on the qubit $b1$, which enlarges the neighborhood of $v_1$. The next step shows the Pauli $Y$ measurement on $v_1$, which establishes the edge $(b_1,v_2)$, necessary for the next measurement, as well as the edge $(r_1,r_2)$, necessary for restoring the cluster state, and it removes the qubit $r_1$ from the measurement collective, which removes the necessity to measure it in Pauli $Z$ basis. The last step shows the final local complementation on $b_1$, which restores a similar configuration as in the initial step, merely $r_1$ is replaced by $r_2$ and the exclusive neighbors switch from $b_1$ to $v_2$ and from $v_1$ to $b_1$.}
    \label{fig:zipperExplanation}
\end{figure*}

In this section we show that the zipper scheme restores the structure of the remaining cluster state while creating a Bell state between the two end qubits. We use the graphical and set theoretical rules for Pauli measurements and local Clifford operations from the Refs. \cite{HeinPRA2004,heinEntanglementGraphstates}. We consider a 2D cluster state where we apply the zipper scheme along a diagonal path of qubits, labeled from $v_1$ to $v_n$, to generate a Bell state between the two end qubits $b_1$ and $b_2$ of the path, see also Fig.~\ref{fig:zipperExplanation}. The measurements on the diagonal impact the neighborhood along the path. We distinguish two kinds of qubits in the vicinity of the measured qubits, namely exclusive neighbor qubits that are in the neighborhood of only one measured or the Bell state qubit, labeled with $e_i$ variables, and qubits that are neighbors of at least two measured qubits on the diagonal, labeled with $r_i$ variables. The left-most sub-figure in Fig.~\ref{fig:zipperExplanation} sketches the situation for a diagonal consisting of six qubits. In the following we explain the zipper scheme step by step by starting with the Pauli $X$ measurement of $v_1$.

We recall from section \ref{app:sec:LCUPM} that a Pauli $X$ measurement on the qubit $v_1$ corresponds to a local complementation on a special neighbor (which we choose to be $b_1 \in N_{v_{1}}$), followed by a Pauli $Y$ measurement on the qubit $v_1$ and another local complementation on the special neighbor.

When measuring $v_1$ the first step corresponds to a local complementation on the selected, special neighbor, namely on qubit $b_1$. This extends the edge set of the subgraph induced by the neighbors of $b_1$ to a fully-connected subgraph on the vertex set $\{e_1,e_2,r_1,v_1\}$, see the second sub-figure in Fig.~\ref{fig:zipperExplanation}. Note that this also extends the neighborhood of the measurement qubit $v_1$, which we virtually measure in the Pauli $Y$ basis next. We note that this virtual measurement corresponds to a local complementation on $v_1$ followed by a Pauli $Z$ measurement. The virtual Pauli $Y$ measurement induces many changes to the edge set of the graph, as one can see in the third sub-figure in Fig.~\ref{fig:zipperExplanation}. For example, the qubit $r_1$ which was a neighbor of two qubits along the measurement path gets detached from both qubits $b_1$ and $v_2$ on the path. This implies that $r_1$ is not a direct neighbor neither to the qubits on the diagonal path nor the Bell state qubits. Furthermore, this circumstance removes the necessity to measure it in the Pauli $Z$ basis. Also, the virtual Pauli $Y$ measurement creates edges between the qubits $r_1$ and $r_2$ as well as between $b_1$ and $v_2$, where the latter is the key ingredient to continue with the X protocol. Finally, note that exclusive neighbors hop from $b_1$ to $v_2$ and from $v_1$ to $b_1$. The final step is another local complementation on the special neighbor $b_1$, which removes the fully-connected graph of the vertices $\{v_2,r_2,e_3\}$ from the graph state, see the last sub-figure in Fig.~\ref{fig:zipperExplanation}. This last step removes the edge between the next qubit to measure $v_2$ and $r_2$, and thus, it restores a similar setting as for measuring the qubit $v_1$. In this setting we find that $r_2$ is a neighbor of the Bell state qubit $b_1$ and $v_3$ in a similar manner as $r_1$ was for $b_1$ and $v_2$ in the previous step. Specifically, since $r_2$ is now in a similar situation as $r_1$ was initially the same observations outlined above apply when measuring $v_2$. In summary, when we measure the qubits $v_1$ to $v_n$, $b_1$ connects step-by step to the next qubit until it ends up connecting to $b_2$. The exclusive neighbors $e_i$ toggle between $b_1$ and the qubit to be measured $v_j$ in each measurement step $j$. Importantly, all qubits $r_i$ detach from the measurement path and connect with each other in such a way that they restore the underlying cluster state structure. The exclusive neighbors $e_i$ accompany the Bell states and measurement path until the very end which results in the necessity to remove them in the end.

\section{Horizontal and vertical transport of multiple data lines in a 2D cluster state}\label{sec:appendix:parallel}
\begin{figure}[ht]
    \centering
        \includegraphics[trim={0cm 11cm 0 0},clip]{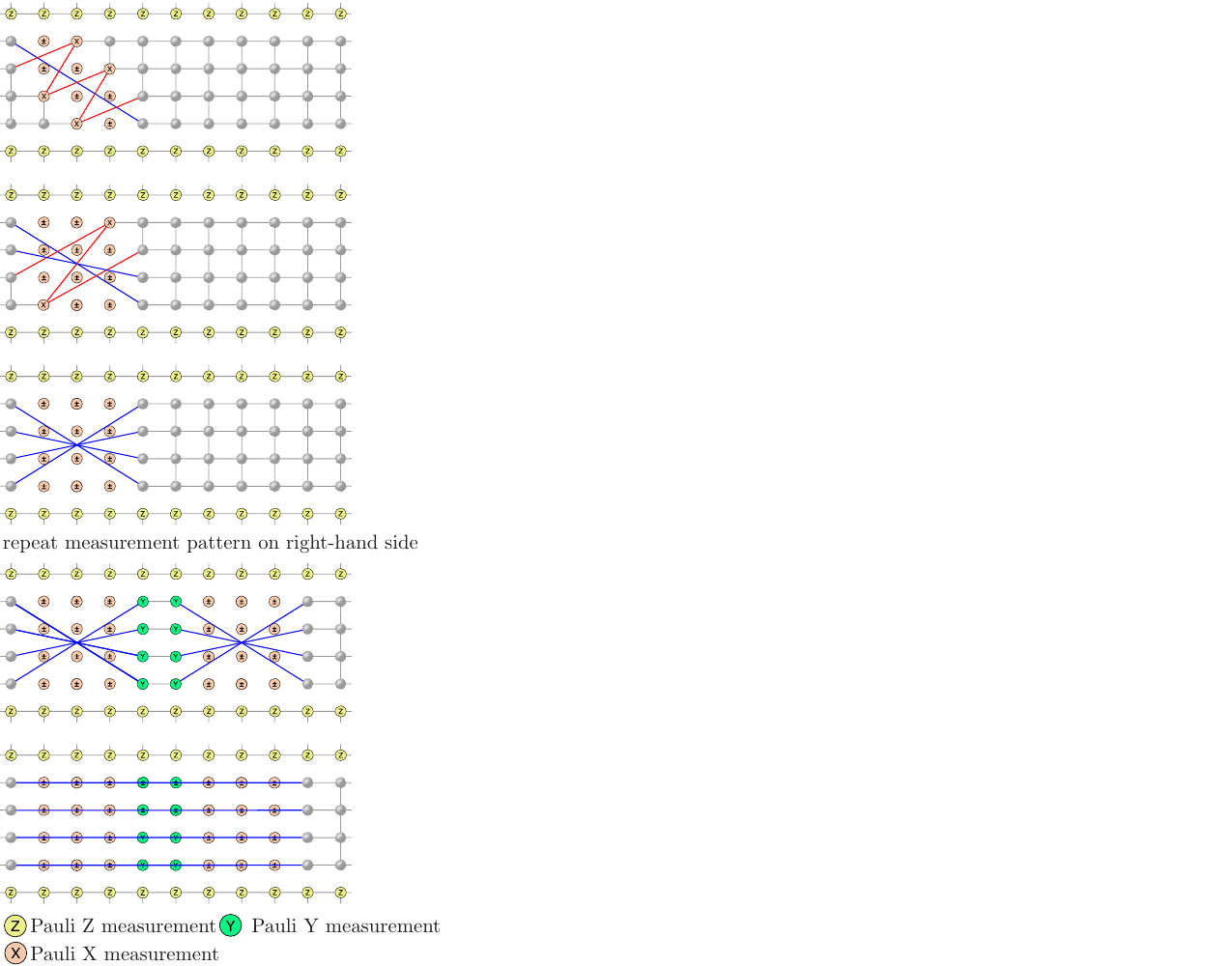}
    \caption{Measurement pattern for a four qubit parallel transport in horizontal direction. The first three sub-figures show the measurement pattern used to achieve an inversion of order of the input data. The fourth sub-figure repeats the measurement strategy a second time to revert this inversion. The last figure shows the final result after performing the Pauli $Y$ measurements on the intermediate qubits.}
    \label{fig:ParallelFourQubits}
\end{figure}

In this section we show how to transport $n$ parallel Bell states $(A_1,B_1),\dots, (A_n,B_n)$ on a straight horizontal or vertical line in a 2D cluster state. To implement such a parallel transport, we first apply the zipper scheme to diagonally transport the first data line ($A_1$) to the last line ($A_n$). By using the cluster state restoring property of the zipper scheme, we transport in a second step the second data line ($A_2$) to the second last position ($A_{n-1}$) again via the zipper scheme. We repeat this until we reversed the order of all lines, namely to $(A_n,A_{n-1},\dots A_1)$. Next, we apply the same measurement strategy again. We observe that we recover the original order of the lines through that. However, we note that we must keep one qubit on each permuted line, resulting in $n$ qubits between the two measurement strategies, to have cluster states before starting the second application of the zipper scheme. The final step is to apply Pauli $Y$ measurements to merge these intermediate qubits on each line in between to the create the final Bell states. If we have a distance of $N$ qubits between a Bell pair, we can retrieve from a 2D cluster state at most $\mathcal{O}(N/2)$ Bell states, and note that a fixed number of measurements, scaling with the number of parallel lines, is required to transport parallel lines. In Fig.~\ref{fig:ParallelFourQubits} we show the measurement strategy to transport four Bell states in parallel step by step. 
Notice that in contrast with using the isolation strategy, there are no disconnected lines between data lines. This implies (roughly) a factor of two improvement in required resources to transport multiple data lines in parallel. 

\section{Straight to diagonal routing of multiple data lines in a 2D cluster state}\label{sec:appendix:straight_to_diagonal}
\begin{figure}[ht]
    \centering
\includegraphics[trim={0cm 18.6cm 0 0},clip]{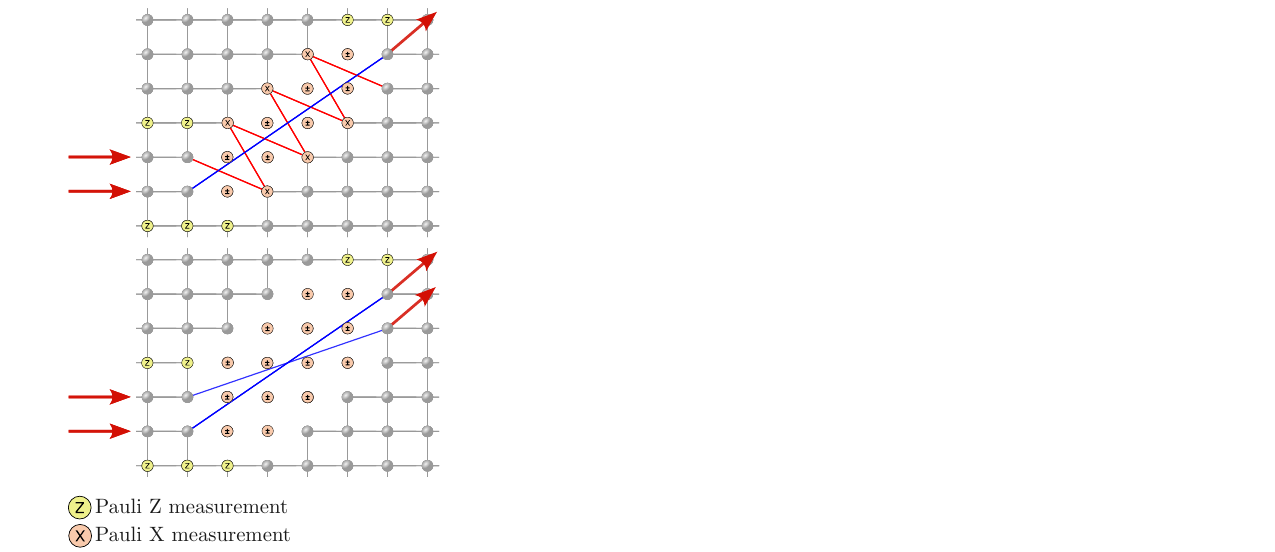}
\caption{Measurement pattern to change between parallel and diagonal information transport is shown. The qubit in the bottom of the parallel line stack has been transported first by  with the zipper scheme. The next qubit in the stack is transported by applying the zipper scheme along the seam, produced by the first application of the zipper scheme.}
    \label{fig:parallel2diag}
\end{figure}

In this section we present the measurement pattern to change from parallel to diagonal data lines of a quantum data bus in a 2D cluster state. Applying the zipper scheme to the last data line in a parallel transport together with the restoration property of the zipper scheme enables to transport the next qubit along the seam closed by the zipper scheme. The top part of Fig.~\ref{fig:parallel2diag} depicts the result after switching the direction of the last qubit of a horizontal quantum data bus to diagonal (blue) using the zipper scheme. We transport the second-last qubit now by measuring along the seam (red line) produced by the zipper scheme of the last qubit in the first step. The figure at the bottom of Fig.~\ref{fig:parallel2diag} shows the result after the both qubits of the quantum data bus switched the direction from horizontal to diagonal. Note that the length of the initial path for the zipper scheme determines the number of qubits for the transfer, where a length of $2n$ enables to transfer $n$ qubits. Furthermore, we note that this approach inverts the order of data lines in the quantum data bus, similar as for parallel transport (see also Appendix~\ref{sec:appendix:parallel} for more information).

\begin{figure}[h]
    \centering
    \includegraphics[trim={0cm 19.6cm 0 0},clip]{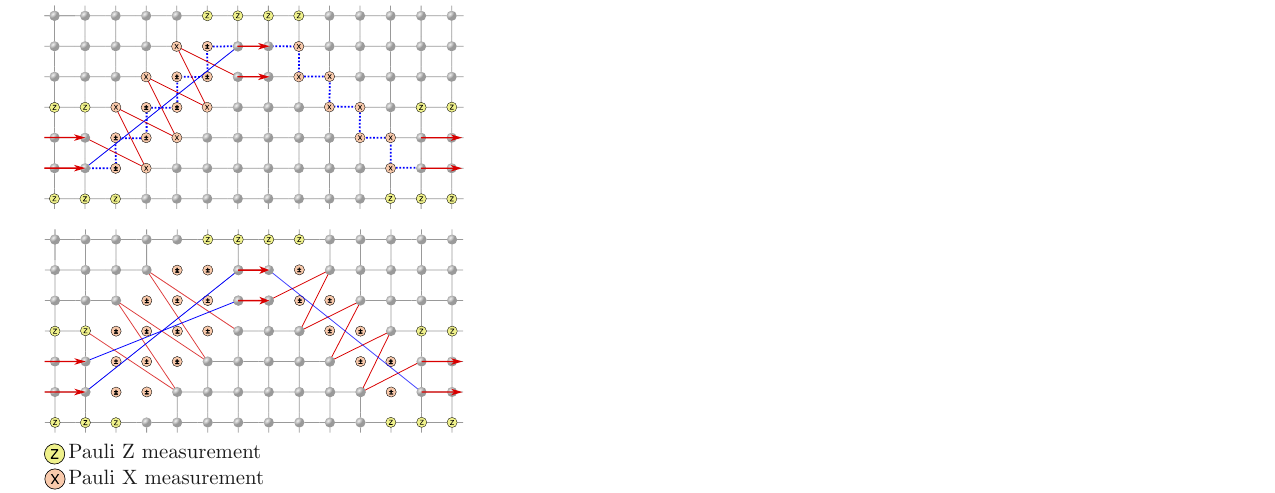}
    \caption{Measurement strategy to perform a \textquotesingle V\textquotesingle -turn on two parallel data lines. Top sketch illustrates the application of the zipper scheme along the blue, dashed path to propagate the bottom data line to the top, and the next data line will be propagated to the top by applying the zipper scheme along the red seam created the zipper scheme. The bottom part sketches the \textquotesingle V\textquotesingle -turn after both data lines have been brought up and the bottom line has been brought down. As a final step the two qubits in the middle have to be merged by Pauli $X$ measurements.}
    \label{fig:VturnParallel}
\end{figure}
\section{\textquotesingle V\textquotesingle -turn of multiple data lines in a 2D cluster state}\label{app:V}
In this section we demonstrate the measurement pattern to perform a \textquotesingle V\textquotesingle -turn of parallel data lines. It basically boils down to applying the same measurement strategy as in the previous section of Appendix~\ref{sec:appendix:straight_to_diagonal}, i.e. to change from straight to diagonal transport, just twice. In the top part of Fig.~\ref{fig:VturnParallel} we applied the zipper scheme along the blue, dashed paths in order to propagate the bottom data line to the top. We use the closed seam (red) of the zipper scheme, like for straight-to-diagonal transport, to propagate the next data line to the top. To propagate the qubits now from top to bottom again, we apply the previous measurement pattern for all data lines again, but now downwards in such a way that the resulting paths correspond to a \textquotesingle V\textquotesingle. We note that we require one intermediate qubit per line at the turning point to enable the full functionality of the zipper scheme. In the bottom part of Fig.~\ref{fig:VturnParallel} the intermediate result for a \textquotesingle V\textquotesingle -turn of two parallel data lines to the top and one to the bottom is shown. To turn around $n$ data lines, a total number of $2(n\times n)$ is required.

\begin{figure}[h]
    \centering
    \includegraphics[trim={0cm 20.5cm 0 0},clip]{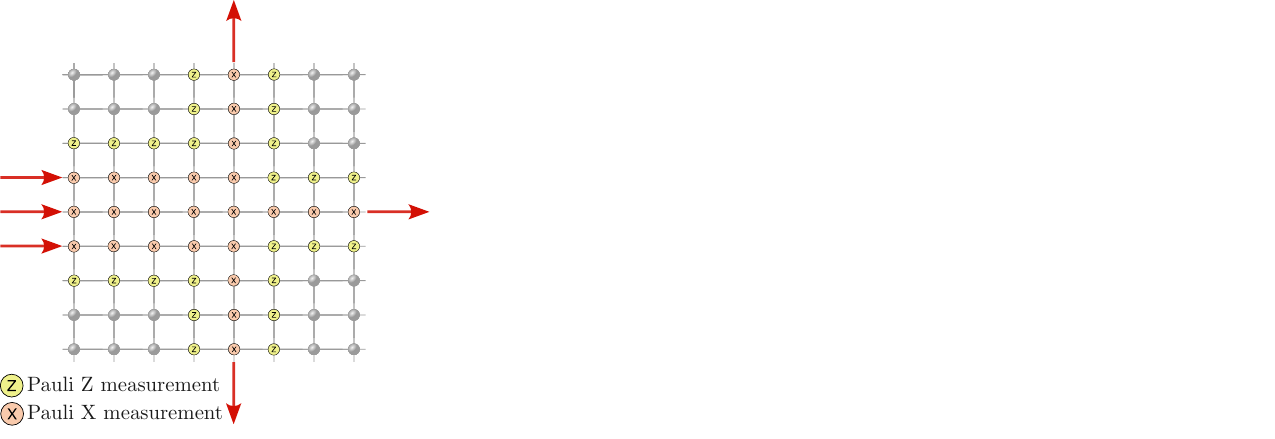}
    \caption{Pauli $Z$ measurements are depicted by the yellow badges, and those cut out the measurement path. By measuring the qubits along the orange lines in the Pauli $X$ basis, we achieve the splitting of the three data lines.}
    \label{fig:fadin}
\end{figure}

\begin{figure*}[!ht]
    \includegraphics[trim={0cm 22.2cm 0 0},clip]{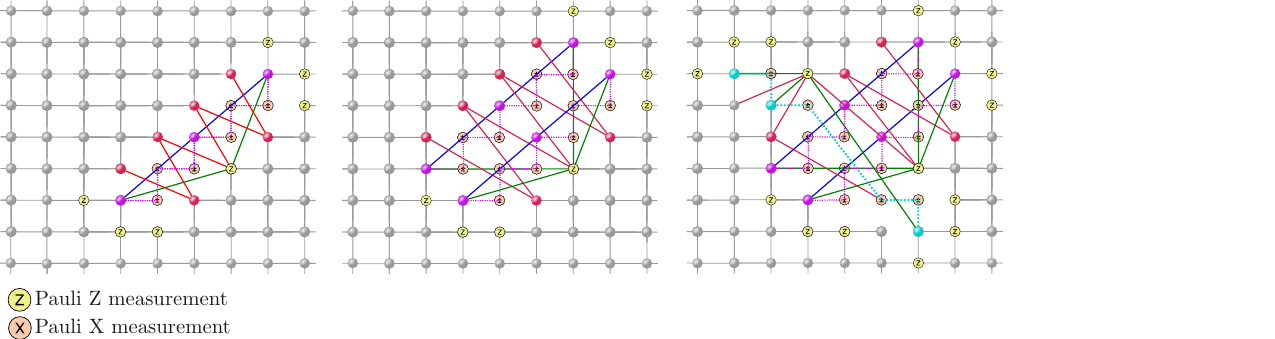}
    \caption{Linear cluster or GHZ extraction: In the first sub-figure we apply the zipper scheme along a (purple, dashed) staircase-shaped path to extract a one-dimensional cluster or GHZ state (purple qubits) from the 2D cluster state by keeping intermediate qubits unmeasured, and measuring additional qubits (yellow) within the Pauli $Z$ basis. In the middle we demonstrate that the remaining cluster state, due to the zipper scheme, enables to extract two one-dimensional cluster or GHZ state adjacent to each other. In the right part an additional one-dimensional cluster or GHZ state (turquoise) is extracted orthogonal to two established states.}
    \label{fig:zipper1Dcluster} 
\end{figure*}

\section{Merging and splitting of  data lines of the data bus} \label{app:merging}
We achieve a merging and splitting of the data lines that constitute the quantum data bus by placing Pauli $Z$ measurements at appropriate positions. In Fig.~\ref{fig:fadin} we show a splitting of a three qubit data line bundle where the yellow measurements correspond to Pauli $Z$ measurements for cutting out the data lines, which are merged subsequently by the orange Pauli $X$ measurements.

\section{Zipper scheme for one-dimensional cluster state and GHZ state extraction}\label{app:ghz}
In this section, we show how to apply the zipper scheme to extract one-dimensional cluster states and GHZ states from a two dimensional cluster state. We point out that by using insights from Ref.~\cite{dejong2023extracting} together with our strategy we generate GHZ states. The simplest way to extract a linear cluster state from a two dimensional cluster state is by applying the zipper scheme and not measuring qubits on the path. In the first part from the left of Fig.~\ref{fig:zipper1Dcluster} we show how to extract a GHZ state or a three qubit linear cluster state from a 2D cluster state by applying the zipper scheme along the purple measurement path. However, we note that this comes with an increased number of qubits we have to measure in the Pauli $Z$ basis (yellow qubit along the seam), because (green) edges to the target state remain in addition to the (red) edges due to the zipper scheme. Overall, the additional Pauli $Z$ measurements of qubits reduce the connectivity in the remaining resource state.

\subsection{Crossing of 1D cluster states and GHZ states}
Choosing the locations of the qubits in a clever way enables us to extract multiple linear cluster states by measuring along a staircase path, similarly as discovered for Bell states. In particular, we split the staircase into two sub-staircase segments  such that the beginning and the ending of each stair segment look in the opposite direction. From that condition follows directly, that half of the staircase qubits cannot be chosen, if we want to measure parallel staircases directly adjacent. Furthermore, if we choose the remaining qubits at the same position for all parallel staircase paths, we only have to remove a single exclusive neighbor for each qubit we keep in the linear cluster state. In the middle part of Fig.~\ref{fig:zipper1Dcluster} two parallel linear cluster or GHZ states of size three are extracted by two sub-staircase measurement paths from a two dimensional cluster state, whereas the yellow qubits need to removed from the target states by Pauli $Z$ measurements.

Moreover, we can still extract further linear cluster or GHZ states orthogonal to first ones from the resulting resource state, but the total number of crossing linear cluster states reduces by the number qubits unmeasured on the path. In the right part of Fig.~\ref{fig:zipper1Dcluster} we show a single linear cluster or GHZ state (turquoise) crossing the two already established ones. Another (yellow) qubits needs to be removed, when we extract the orthogonal linear cluster states.

In summary, using the zipper scheme to distill a linear cluster or GHZ state from the 2D cluster works, but demands a measurement for each additional qubit we add to the linear cluster state. Additionally, the generation of GHZ states \cite{dejong2023extracting} with more than three qubits from the linear cluster state requires approximately half of the qubits to measured. 

\newpage
\bibliography{main}

@Article{Hahn2019,
author={Hahn, F.
and Pappa, A.
and Eisert, J.},
title={Quantum network routing and local complementation},
journal={npj Quantum Information},
year={2019},
month={Sep},
day={06},
volume={5},
number={1},
pages={76},
issn={2056-6387},
doi={10.1038/s41534-019-0191-6},
url={https://doi.org/10.1038/s41534-019-0191-6}
}

@article{HeinPRA2004,
  title = {Multiparty entanglement in graph states},
  author = {Hein, M. and Eisert, J. and Briegel, H. J.},
  journal = {Phys. Rev. A},
  volume = {69},
  issue = {6},
  pages = {062311},
  numpages = {20},
  year = {2004},
  month = {Jun},
  publisher = {American Physical Society},
  doi = {10.1103/PhysRevA.69.062311},
  url = {https://link.aps.org/doi/10.1103/PhysRevA.69.062311}
}

@article{dejong2023extracting,
  title = {Extracting GHZ states from linear cluster states},
  author = {de Jong, J. and Hahn, F. and Tcholtchev, N. and Hauswirth, M. and Pappa, A.},
  journal = {Phys. Rev. Res.},
  volume = {6},
  issue = {1},
  pages = {013330},
  numpages = {7},
  year = {2024},
  month = {Mar},
  publisher = {American Physical Society},
  doi = {10.1103/PhysRevResearch.6.013330},
  url = {https://link.aps.org/doi/10.1103/PhysRevResearch.6.013330}
}

@article{mannalathPRAmultiparty,
  title = {Multiparty entanglement routing in quantum networks},
  author = {Mannalath, Vaisakh and Pathak, Anirban},
  journal = {Phys. Rev. A},
  volume = {108},
  issue = {6},
  pages = {062614},
  numpages = {15},
  year = {2023},
  month = {Dec},
  publisher = {American Physical Society},
  doi = {10.1103/PhysRevA.108.062614},
  url = {https://link.aps.org/doi/10.1103/PhysRevA.108.062614}
}

@article{Pirker_2018,
doi = {10.1088/1367-2630/aac2aa},
url = {https://dx.doi.org/10.1088/1367-2630/aac2aa},
year = {2018},
month = {may},
publisher = {IOP Publishing},
volume = {20},
number = {5},
pages = {053054},
author = {A Pirker and J Wallnöfer and W Dür},
title = {Modular architectures for quantum networks},
journal = {New Journal of Physics}
}

@article{morruiz2023influence,
  author={Mor-Ruiz, Maria Flors and Dür, Wolfgang},
  journal={IEEE Journal on Selected Areas in Communications}, 
  title={Influence of noise in entanglement-based quantum networks}, 
  year={2024},
  volume={},
  number={},
  pages={1-1},
  doi={10.1109/JSAC.2024.3380089}}

@article{MiguelRamiro2023optimizedquantum,
  doi = {10.22331/q-2023-02-09-919},
  url = {https://doi.org/10.22331/q-2023-02-09-919},
  title = {Optimized {Q}uantum {N}etworks},
  author = {Miguel-Ramiro, Jorge and Pirker, Alexander and D{\"{u}}r, Wolfgang},
  journal = {{Quantum}},
  issn = {2521-327X},
  publisher = {{Verein zur F{\"{o}}rderung des Open Access Publizierens in den Quantenwissenschaften}},
  volume = {7},
  pages = {919},
  month = feb,
  year = {2023}
}

@article{Pirker_2019,
doi = {10.1088/1367-2630/ab05f7},
url = {https://dx.doi.org/10.1088/1367-2630/ab05f7},
year = {2019},
month = {mar},
publisher = {IOP Publishing},
volume = {21},
number = {3},
pages = {033003},
author = {A Pirker and W Dür},
title = {A quantum network stack and protocols for reliable entanglement-based networks},
journal = {New Journal of Physics}
}

@ARTICLE{butterfly,
  author={Leung, Debbie and Oppenheim, Jonathan and Winter, Andreas},
  journal={IEEE Transactions on Information Theory}, 
  title={Quantum Network Communication—The Butterfly and Beyond}, 
  year={2010},
  volume={56},
  number={7},
  pages={3478-3490},
  doi={10.1109/TIT.2010.2048442}}

@article{RaussendorfOneWayPRL2001,
  title = {A One-Way Quantum Computer},
  author = {Raussendorf, Robert and Briegel, Hans J.},
  journal = {Phys. Rev. Lett.},
  volume = {86},
  issue = {22},
  pages = {5188--5191},
  numpages = {0},
  year = {2001},
  month = {May},
  publisher = {American Physical Society},
  doi = {10.1103/PhysRevLett.86.5188},
  url = {https://link.aps.org/doi/10.1103/PhysRevLett.86.5188}
}

@article{DahlbergNPcomplete,
author = {Dahlberg, Axel  and Wehner, Stephanie },
title = {Transforming graph states using single-qubit operations},
journal = {Philosophical Transactions of the Royal Society A: Mathematical, Physical and Engineering Sciences},
volume = {376},
number = {2123},
pages = {20170325},
year = {2018},
doi = {10.1098/rsta.2017.0325}
}

@article{QuantumInternetKimble,
	author = {Kimble, H.  J. },
	date = {2008/06/01},
	doi = {10.1038/nature07127},
	id = {Kimble2008},
	isbn = {1476-4687},
	journal = {Nature},
	number = {7198},
	pages = {1023--1030},
	title = {The quantum internet},
	url = {https://doi.org/10.1038/nature07127},
	volume = {453},
	year = {2008}}

@article{
QuantumInternetWehner,
author = {Stephanie Wehner  and David Elkouss  and Ronald Hanson },
title = {Quantum internet: A vision for the road ahead},
journal = {Science},
volume = {362},
number = {6412},
pages = {eaam9288},
year = {2018},
doi = {10.1126/science.aam9288}}

@article{DistributedQuComputing,
  title = {Distributed quantum computation over noisy channels},
  author = {Cirac, J. I. and Ekert, A. K. and Huelga, S. F. and Macchiavello, C.},
  journal = {Phys. Rev. A},
  volume = {59},
  issue = {6},
  pages = {4249--4254},
  numpages = {0},
  year = {1999},
  month = {Jun},
  publisher = {American Physical Society},
  doi = {10.1103/PhysRevA.59.4249},
  url = {https://link.aps.org/doi/10.1103/PhysRevA.59.4249}
}

@article{Das17,
  title = {Robust quantum network architectures and topologies for entanglement distribution},
  author = {Das, Siddhartha and Khatri, Sumeet and Dowling, Jonathan P.},
  journal = {Phys. Rev. A},
  volume = {97},
  issue = {1},
  pages = {012335},
  numpages = {12},
  year = {2018},
  month = {Jan},
  publisher = {American Physical Society},
  doi = {10.1103/PhysRevA.97.012335},
  url = {https://link.aps.org/doi/10.1103/PhysRevA.97.012335}
}

@Article{Pirandola2019,
author={Pirandola, Stefano},
title={End-to-end capacities of a quantum communication network},
journal={Communications Physics},
year={2019},
month={May},
day={17},
volume={2},
number={1},
pages={51},
issn={2399-3650},
doi={10.1038/s42005-019-0147-3},
url={https://doi.org/10.1038/s42005-019-0147-3}
}

@article{Gyongyosi18,
  title = {Decentralized base-graph routing for the quantum internet},
  author = {Gyongyosi, Laszlo and Imre, Sandor},
  journal = {Phys. Rev. A},
  volume = {98},
  issue = {2},
  pages = {022310},
  numpages = {9},
  year = {2018},
  month = {Aug},
  publisher = {American Physical Society},
  doi = {10.1103/PhysRevA.98.022310},
  url = {https://link.aps.org/doi/10.1103/PhysRevA.98.022310}
}

@Article{Gyongyosi2017,
author={Gyongyosi, Laszlo
and Imre, Sandor},
title={Entanglement-Gradient Routing for Quantum Networks},
journal={Scientific Reports},
year={2017},
volume={7},
number={1},
pages={14255},
issn={2045-2322},
doi={10.1038/s41598-017-14394-w},
url={https://doi.org/10.1038/s41598-017-14394-w}
}

@ARTICLE{Caleffi17, 
author={M. Caleffi}, 
journal={IEEE Access}, 
title={Optimal Routing for Quantum Networks}, 
year={2017}, 
volume={5}, 
number={}, 
pages={22299-22312},  
doi={10.1109/ACCESS.2017.2763325}, 
ISSN={2169-3536}, 
month={},}

@article{Schoute2016,
	author = "Schoute, Eddie and Mancinska, Laura and Islam, Tanvirul and Kerenidis, Iordanis and Wehner, Stephanie",
	journal = "arXiv preprint arXiv:1610.05238",
	title = "{Shortcuts to quantum network routing}",
	url = "http://arxiv.org/abs/1610.05238",
	year = "2016"
}

@article{Meter2013b,
	author = "{Van Meter}, Rodney and Satoh, Takahiko and Ladd, Thaddeus D. and Munro, William J. and Nemoto, Kae",
	day = "01",
	doi = "10.1007/s13119-013-0026-2",
	issn = "2076-0329",
	journal = "Networking Science",
	month = "Dec",
	number = "1",
	pages = "82--95",
	title = "{Path selection for quantum repeater networks}",
	url = "http://dx.doi.org/10.1007/s13119-013-0026-2",
	volume = "3",
	year = "2013"
}

@ARTICLE{Cacciapuoti18,
       author={Cacciapuoti, Angela Sara and Caleffi, Marcello and Tafuri, Francesco and Cataliotti, Francesco Saverio and Gherardini, Stefano and Bianchi, Giuseppe},
  journal={IEEE Network}, 
  title={Quantum Internet: Networking Challenges in Distributed Quantum Computing}, 
  year={2020},
  volume={34},
  number={1},
  pages={137-143},
  doi={10.1109/MNET.001.1900092}
}

@article{Markham08,
	author = "Markham, Damian and Sanders, Barry C.",
	doi = "10.1103/PhysRevA.78.042309",
	issue = "4",
	journal = "Phys. Rev. A",
	month = "Oct",
	numpages = "17",
	pages = "042309",
	publisher = "American Physical Society",
	title = "{Graph states for quantum secret sharing}",
	url = "https://link.aps.org/doi/10.1103/PhysRevA.78.042309",
	volume = "78",
	year = "2008"
}

@article{Hillery99,
	author = "Hillery, Mark and {Bu\ifmmode \check{z}\else \v{z}\fi{}ek}, Vladim{\'i}r and Berthiaume, Andr{\'e}",
	doi = "10.1103/PhysRevA.59.1829",
	issue = "3",
	journal = "Phys. Rev. A",
	month = "Mar",
	numpages = "0",
	pages = "1829--1834",
	publisher = "American Physical Society",
	title = "{Quantum secret sharing}",
	url = "https://link.aps.org/doi/10.1103/PhysRevA.59.1829",
	volume = "59",
	year = "1999"
}

@Article{Wallnofer2019,
author={Walln{\"o}fer, J.
and Pirker, A.
and Zwerger, M.
and D{\"u}r, W.},
title={Multipartite state generation in quantum networks with optimal scaling},
journal={Scientific Reports},
year={2019},
month={Jan},
day={22},
volume={9},
number={1},
pages={314},
issn={2045-2322},
doi={10.1038/s41598-018-36543-5},
url={https://doi.org/10.1038/s41598-018-36543-5}
}

@article{Meignant2019,
  title = {Distributing graph states over arbitrary quantum networks},
  author = {Meignant, Cl\'ement and Markham, Damian and Grosshans, Fr\'ed\'eric},
  journal = {Phys. Rev. A},
  volume = {100},
  issue = {5},
  pages = {052333},
  numpages = {6},
  year = {2019},
  month = {Nov},
  publisher = {American Physical Society},
  doi = {10.1103/PhysRevA.100.052333},
  url = {https://link.aps.org/doi/10.1103/PhysRevA.100.052333}
}

@article{Cuquet2012,
  title = {Growth of graph states in quantum networks},
  author = {Cuquet, Mart\'{\i} and Calsamiglia, John},
  journal = {Phys. Rev. A},
  volume = {86},
  issue = {4},
  pages = {042304},
  numpages = {15},
  year = {2012},
  month = {Oct},
  publisher = {American Physical Society},
  doi = {10.1103/PhysRevA.86.042304},
  url = {https://link.aps.org/doi/10.1103/PhysRevA.86.042304}
}

@article{Epping_2016,
doi = {10.1088/1367-2630/18/5/053036},
url = {https://dx.doi.org/10.1088/1367-2630/18/5/053036},
year = {2016},
month = {may},
publisher = {IOP Publishing},
volume = {18},
number = {5},
pages = {053036},
author = {Michael Epping and Hermann Kampermann and Dagmar Bruß},
title = {Large-scale quantum networks based on graphs},
journal = {New Journal of Physics}
}

@article{Eldredge2018,
  title = {Optimal and secure measurement protocols for quantum sensor networks},
  author = {Eldredge, Zachary and Foss-Feig, Michael and Gross, Jonathan A. and Rolston, S. L. and Gorshkov, Alexey V.},
  journal = {Phys. Rev. A},
  volume = {97},
  issue = {4},
  pages = {042337},
  numpages = {9},
  year = {2018},
  month = {Apr},
  publisher = {American Physical Society},
  doi = {10.1103/PhysRevA.97.042337},
  url = {https://link.aps.org/doi/10.1103/PhysRevA.97.042337}
}

@article{BennettTele,
  title = {Teleporting an unknown quantum state via dual classical and Einstein-Podolsky-Rosen channels},
  author = {Bennett, Charles H. and Brassard, Gilles and Cr\'epeau, Claude and Jozsa, Richard and Peres, Asher and Wootters, William K.},
  journal = {Phys. Rev. Lett.},
  volume = {70},
  issue = {13},
  pages = {1895--1899},
  numpages = {0},
  year = {1993},
  month = {Mar},
  publisher = {American Physical Society},
  doi = {10.1103/PhysRevLett.70.1895},
  url = {https://link.aps.org/doi/10.1103/PhysRevLett.70.1895}
}

@article{Azuma2021,
    author = {Azuma, Koji and Bäuml, Stefan and Coopmans, Tim and Elkouss, David and Li, Boxi},
    title = "{Tools for quantum network design}",
    journal = {AVS Quantum Science},
    volume = {3},
    number = {1},
    pages = {014101},
    year = {2021},
    month = {02},
    issn = {2639-0213},
    doi = {10.1116/5.0024062}
}

@article{Azuma2023,
  title = {Quantum repeaters: From quantum networks to the quantum internet},
  author = {Azuma, Koji and Economou, Sophia E. and Elkouss, David and Hilaire, Paul and Jiang, Liang and Lo, Hoi-Kwong and Tzitrin, Ilan},
  journal = {Rev. Mod. Phys.},
  volume = {95},
  issue = {4},
  pages = {045006},
  numpages = {66},
  year = {2023},
  month = {Dec},
  publisher = {American Physical Society},
  doi = {10.1103/RevModPhys.95.045006},
  url = {https://link.aps.org/doi/10.1103/RevModPhys.95.045006}
}

@article{Illiano2022,
title = {Quantum Internet protocol stack: A comprehensive survey},
journal = {Computer Networks},
volume = {213},
pages = {109092},
year = {2022},
issn = {1389-1286},
doi = {https://doi.org/10.1016/j.comnet.2022.109092},
url = {https://www.sciencedirect.com/science/article/pii/S1389128622002250},
author = {Jessica Illiano and Marcello Caleffi and Antonio Manzalini and Angela Sara Cacciapuoti}
}

@ARTICLE{Cacciapuoti2020,
  author={Cacciapuoti, Angela Sara and Caleffi, Marcello and Van Meter, Rodney and Hanzo, Lajos},
  journal={IEEE Transactions on Communications}, 
  title={When Entanglement Meets Classical Communications: Quantum Teleportation for the Quantum Internet}, 
  year={2020},
  volume={68},
  number={6},
  pages={3808-3833},
  doi={10.1109/TCOMM.2020.2978071}}

@article{Avis2023,
  title = {Analysis of multipartite entanglement distribution using a central quantum-network node},
  author = {Avis, Guus and Rozp\ifmmode \mbox{\k{e}}\else \k{e}\fi{}dek, Filip and Wehner, Stephanie},
  journal = {Phys. Rev. A},
  volume = {107},
  issue = {1},
  pages = {012609},
  numpages = {36},
  year = {2023},
  month = {Jan},
  publisher = {American Physical Society},
  doi = {10.1103/PhysRevA.107.012609},
  url = {https://link.aps.org/doi/10.1103/PhysRevA.107.012609}
}

@article{Danos2006,
  title = {Determinism in the one-way model},
  author = {Danos, Vincent and Kashefi, Elham},
  journal = {Phys. Rev. A},
  volume = {74},
  issue = {5},
  pages = {052310},
  numpages = {6},
  year = {2006},
  month = {Nov},
  publisher = {American Physical Society},
  doi = {10.1103/PhysRevA.74.052310},
  url = {https://link.aps.org/doi/10.1103/PhysRevA.74.052310}
}

@article{Browne_2007,
doi = {10.1088/1367-2630/9/8/250},
url = {https://dx.doi.org/10.1088/1367-2630/9/8/250},
year = {2007},
month = {aug},
publisher = {},
volume = {9},
number = {8},
pages = {250},
author = {Daniel E Browne and Elham Kashefi and Mehdi Mhalla and Simon Perdrix},
title = {Generalized flow and determinism in measurement-based quantum computation},
journal = {New Journal of Physics}
}

@InProceedings{heinEntanglementGraphstates,
    author       = {Hein, M. and D{\"u}r, W. and Eisert, J. and Raussendorf, R. and Van den Nest, M. and Briegel, H.-J.},
    title        = {Entanglement in graph states and its applications},
    booktitle    = {Quantum Computers,  Algorithms and Chaos},
    year         = {2006},
    editor       = {Casati, G. and Shepelyansky, D. L. and Zoller, P. and Benenti, G.},
    volume       = {162},
    series       = {Proceedings of the International School of Physics "Enrico Fermi"},
    pages        = {115–218},
    publisher    = {IOS Press},
    url          = {https://doi.org/10.3254/978-1-61499-018-5-115},
    doi          = {10.3254/978-1-61499-018-5-115}
  }

@article{BriegelPRL2001Cluster,
  title = {Persistent Entanglement in Arrays of Interacting Particles},
  author = {Briegel, Hans J. and Raussendorf, Robert},
  journal = {Phys. Rev. Lett.},
  volume = {86},
  issue = {5},
  pages = {910--913},
  numpages = {0},
  year = {2001},
  month = {Jan},
  publisher = {American Physical Society},
  doi = {10.1103/PhysRevLett.86.910},
  url = {https://link.aps.org/doi/10.1103/PhysRevLett.86.910}
}

@article{RaussendorfPRA2003MBQCCluster,
  title = {Measurement-based quantum computation on cluster states},
  author = {Raussendorf, Robert and Browne, Daniel E. and Briegel, Hans J.},
  journal = {Phys. Rev. A},
  volume = {68},
  issue = {2},
  pages = {022312},
  numpages = {32},
  year = {2003},
  month = {Aug},
  publisher = {American Physical Society},
  doi = {10.1103/PhysRevA.68.022312},
  url = {https://link.aps.org/doi/10.1103/PhysRevA.68.022312}
}

@Article{Mantri2017,
author={Mantri, Atul
and Demarie, Tommaso F.
and Fitzsimons, Joseph F.},
title={Universality of quantum computation with cluster states and (X, Y)-plane measurements},
journal={Scientific Reports},
year={2017},
month={Feb},
day={20},
volume={7},
number={1},
pages={42861},
issn={2045-2322},
doi={10.1038/srep42861},
url={https://doi.org/10.1038/srep42861}
}

@misc{morruiz2024imperfect,
      title={Imperfect quantum networks with tailored resource states}, 
      author={Maria Flors Mor-Ruiz and Julius Wallnöfer and Wolfgang Dür},
      year={2024},
      eprint={2403.19778},
      archivePrefix={arXiv},
      primaryClass={quant-ph}
}

@article{SekatskiPRR2020,
  title = {Optimal distributed sensing in noisy environments},
  author = {Sekatski, P. and W\"olk, S. and D\"ur, W.},
  journal = {Phys. Rev. Res.},
  volume = {2},
  issue = {2},
  pages = {023052},
  numpages = {8},
  year = {2020},
  month = {Apr},
  publisher = {American Physical Society},
  doi = {10.1103/PhysRevResearch.2.023052},
  url = {https://link.aps.org/doi/10.1103/PhysRevResearch.2.023052}
}

@article{GisinRevMP2002,
  title = {Quantum cryptography},
  author = {Gisin, Nicolas and Ribordy, Gr\'egoire and Tittel, Wolfgang and Zbinden, Hugo},
  journal = {Rev. Mod. Phys.},
  volume = {74},
  issue = {1},
  pages = {145--195},
  numpages = {0},
  year = {2002},
  month = {Mar},
  publisher = {American Physical Society},
  doi = {10.1103/RevModPhys.74.145},
  url = {https://link.aps.org/doi/10.1103/RevModPhys.74.145}
}

@Article{Briegel2009,
author={Briegel, H. J.
and Browne, D. E.
and D{\"u}r, W.
and Raussendorf, R.
and Van den Nest, M.},
title={Measurement-based quantum computation},
journal={Nature Physics},
year={2009},
month={Jan},
day={01},
volume={5},
number={1},
pages={19-26},
issn={1745-2481},
doi={10.1038/nphys1157},
url={https://doi.org/10.1038/nphys1157}
}

@article{Wallnofer2016,
  title = {Two-dimensional quantum repeaters},
  author = {Walln\"ofer, J. and Zwerger, M. and Muschik, C. and Sangouard, N. and D\"ur, W.},
  journal = {Phys. Rev. A},
  volume = {94},
  issue = {5},
  pages = {052307},
  numpages = {12},
  year = {2016},
  month = {Nov},
  publisher = {American Physical Society},
  doi = {10.1103/PhysRevA.94.052307},
  url = {https://link.aps.org/doi/10.1103/PhysRevA.94.052307}
}

@INPROCEEDINGS{Chen2023,
  author={Chen, Si-Yi and Cacciapuoti, Angela Sara and Chen, Xiu-Bo and Caleffi, Marcello},
  booktitle={ICC 2023 - IEEE International Conference on Communications}, 
  title={Multipartite Entanglement for the Quantum Internet}, 
  year={2023},
  volume={},
  number={},
  pages={3504-3509},
  doi={10.1109/ICC45041.2023.10278645}
}

@INPROCEEDINGS{Viscardi2023MarkovProcess,
  author={Viscardi, Michele and Illiano, Jessica and Cacciapuoti, Angela Sara and Caleffi, Marcello},
  booktitle={2023 IEEE International Conference on Quantum Computing and Engineering (QCE)}, 
  title={Entanglement Distribution in the Quantum Internet: An Optimal Decision Problem Formulation}, 
  year={2023},
  volume={01},
  number={},
  pages={1114-1119},
  doi={10.1109/QCE57702.2023.00126}}

@article{DistQuantumComputingBuhrman1997PRA,
  title = {Substituting quantum entanglement for communication},
  author = {Cleve, Richard and Buhrman, Harry},
  journal = {Phys. Rev. A},
  volume = {56},
  issue = {2},
  pages = {1201--1204},
  numpages = {0},
  year = {1997},
  month = {Aug},
  publisher = {American Physical Society},
  doi = {10.1103/PhysRevA.56.1201},
  url = {https://link.aps.org/doi/10.1103/PhysRevA.56.1201}
}

@misc{caleffi2022distributed,
      title={Distributed Quantum Computing: a Survey}, 
      author={Marcello Caleffi and Michele Amoretti and Davide Ferrari and Daniele Cuomo and Jessica Illiano and Antonio Manzalini and Angela Sara Cacciapuoti},
      year={2022},
      eprint={2212.10609},
      archivePrefix={arXiv},
      primaryClass={quant-ph}
}

@article{Liorni_2021,
doi = {10.1088/1367-2630/abfa63},
url = {https://dx.doi.org/10.1088/1367-2630/abfa63},
year = {2021},
month = {may},
publisher = {IOP Publishing},
volume = {23},
number = {5},
pages = {053021},
author = {Carlo Liorni and Hermann Kampermann and Dagmar Bruß},
title = {Quantum repeaters in space},
journal = {New Journal of Physics}
}

@ARTICLE{LIIEEE2021,
  author={Li, Boxi and Coopmans, Tim and Elkouss, David},
  journal={IEEE Transactions on Quantum Engineering}, 
  title={Efficient Optimization of Cutoffs in Quantum Repeater Chains}, 
  year={2021},
  volume={2},
  number={},
  pages={1-15},
  doi={10.1109/TQE.2021.3099003}}

\end{document}